\documentstyle[11pt,paspconf,epsfig]{article}

\begin{document}

\title{A {\sc{robust}} method for measuring the Hubble parameter}

\author{M.~A. Hendry and S. Rauzy}
\affil{Department of Physics and Astronomy, University of Glasgow,
Glasgow G12 8QQ, UK}

\begin{abstract}
We obtain a robust, non-parametric, estimate of the Hubble constant from
galaxy linear diameters calibrated using HST Cepheid distances. 
Our method is independent of the parametric form of the diameter function
and the spatial distribution of galaxies and is insensitive to Malmquist bias.
We include information on the galaxy rotation velocities; unlike 
Tully-Fisher, however, we retain a fully non-parametric treatment.
We find $H_0=66\pm6$ kms$^{-1}$ Mpc$^{-1}$, somewhat larger than
previous results using galaxy diameters.
\end{abstract}


\section{Introduction}

Despite the recent emergence of a broad consensus in estimates of $H_0$,
the issue of observational selection effects remains an important one for
studies of the distance scale and peculiar velocity field.
Current recipes for eliminating Malmquist bias make parametric model 
assumptions about the distribution function of the distance indicator 
and the selection effects, and the spatial distribution of the observed 
galaxies. There is clearly an advantage in developing more robust 
techniques in 
which only minimal model assumptions are required. In this paper we present 
such a robust method.

\section{Method and Application Using Galaxy Diameters}

Our technique is based on the $C^-$ method of Lynden-Bell (1971), and provides 
an estimate of the cumulative distribution function (CDF) of galaxy diameter
independent of any model assumption about its parametric form. Moreover, the
method may be applied to data of arbitrary spatial distribution and thus requires
no correction for Malmquist bias. The method is, however, applicable only to 
samples which are strictly complete to a given apparent magnitude or diameter. 
We have developed an objective test of the validity of this assumption, based on the 
approach of Efron \& Petrosian (1992). (See Rauzy \& Hendry in prep. for more details).

We applied the method to reconstruct the CDF of linear isophotal diameter, $D$, from
a sample of 4005 galaxies -- complete to an angular diameter limit of $D \geq 1.5'$ -- 
from the LEDA database (Paturel et al. 1997). We carried out our analysis 
using the variables, $m$ and $M$, analogous to apparent magnitude and absolute magnitude,
given by
\begin{equation}
m = 20 - 5 \log_{10} D
\end{equation}
\begin{equation}
M = m - Z = 20 - 5 \log_{10} D - 5 \log_{10} \frac{cz}{H_0} - 25
\end{equation}

We compared the CDF of $M$ from the LEDA galaxies
with that obtained from a set of 14 local calibrators,
with HST Cepheid distances, from Theureau et al. (1997). We then varied
$H_0$ in eq. (2) and, for each value of $H_0$, determined the Kolmogorov-Smirnoff 
(KS) distance between the two CDFs as a function of $M$. We took as our 
`best-fit' estimate of $H_0$ the value which gave the
{\em{minimum\/}} KS distance -- obtaining
$H_0 = 42$ kms$^{-1}$ Mpc$^{-1}$. This value agrees with Sandage (1993a,b), who
used M31 and M101 as standard rulers, and is consistent with the analysis of
Goodwin, Gribbin \& Hendry (1997), who obtained 
$H_0 = 52 \pm 6 \pm 8$ kms$^{-1}$ Mpc$^{-1}$ using galaxy linear diameters 
and a similar calibrating sample. However, their second uncertainty
was an estimate of the difference in the mean
intrinsic diameter of the local calibrating galaxies compared with
the distant sample (even after correction for Malmquist bias) -- a difference
which might be systematically {\em{negative}\/} given the strategy of the HST 
Key Project to observe `Grand Design' spirals (Kennicutt et al. 1995).
We find evidence for a similar negative bias in our calibrating sample: 
galaxies of small diameter are relatively under-represented in the CDF of the local
calibrators. This would lead to a systematic underestimate in the value of $H_0$.

\section{Including Tully-Fisher information}

We can improve our analysis by introducing galaxy rotation velocity to reduce
the dispersion of the distance indicator. This is analogous to the conventional 
Tully-Fisher relation but -- crucially -- retains completely the robustness of our 
previous analysis. In a similar manner to the `Sosie' method (c.f. Paturel et al. 
1998), we select, for each local calibrator, the subset of galaxies from
the distant sample with similar log rotation velocity and morphological type.
For each subset we then reconstruct the diameter function,
assuming initially a fiducial value of $H_0$. Finally we determine the 
value of $H_0$ required to match the {\em{median}\/} of the CDF to the 
observed diameter for that calibrator.

We applied this technique to the KLUN sample of spiral galaxies
(c.f. Theureau et al. 1997). Fig. 1 shows the CDFs reconstructed from the
subsets corresponding to each local calibrator. The median value of the
reconstructed distribution is indicated on 9 of the panels.
The remaining 3 panels correspond to the 4 local calibrators with the
smallest rotation velocities: N300, N598, N925 and N4496A. We exclude these
calibrators since it is not clear from Fig. 1 whether their corresponding
reconstructed CDFs are completely sampled -- due to the presence of the
lower diameter limit. Matching the median values
from the 9 remaining panels to the linear diameters of the 10 remaining
calibrators, and taking the logarithmic mean of these individual values gives
\begin{equation}
H_0 = 66 \pm 6 \, \, {\rm{kms}}^{-1} \, {\rm{Mpc}}^{-1}
\end{equation}

\begin{figure*}
\centerline{\epsfig{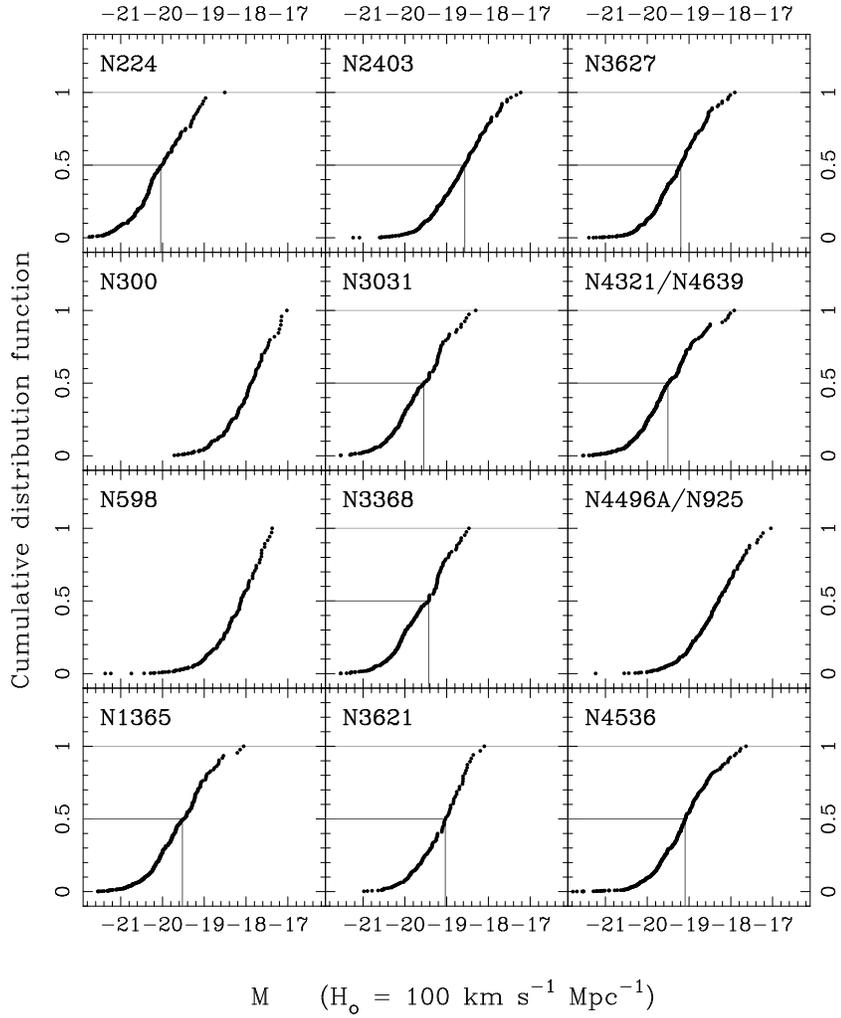}}
\caption{Reconstructed CDFs for subsets of galaxies selected from the KLUN
sample (c.f. Theureau et al. 1997) with similar rotation velocities and
morphological types to the indicated local calibrators.}
\end{figure*}

\section{Discussion}

When we incorporate information on the rotation velocities of the local
calibrators our results are in excellent agreement with recent determinations of 
$H_0$ from the conventional Tully-Fisher relation (c.f. Giovanelli et al 1997).
Note, however, that our analysis is completely free of assumptions about the
form of the galaxy diameter and luminosity function and the conditional
distribution function of diameter at a given value of
$\log V_m$. We do {\bf{not}},
for example, require to assume that this conditional distribution is
Gaussian, nor indeed even that it has zero mean or constant dispersion --
as is often assumed in calibrating the Tully-Fisher relation. In particular,
therefore, we do not require that the Tully-Fisher relation is a straight line,
nor that the distribution of residuals is symmetrical. Our method would remain
applicable if, for example, the distribution of Tully-Fisher residuals
displayed a long `tail' for galaxies of small rotation velocity. Our method
is also completely independent of Malmquist bias corrections, so that our
results are unaffected by the precise `recipe' adopted to correct for
Malmquist bias.

In summary, the robustness and assumption-free nature of this method has
important ramifications for any current debate on the cosmic distance scale.
There appears to be little possibility of any remaining systematic error
in $H_0$ which is introduced at the stage of linking the primary and
secondary distance scales. Improving the calibration of the Cepheid distance
scale, via the lowest rungs of the Cosmic Distance Ladder, must now be the
main priority for the Distance Scale community.

\end{document}